\documentclass[mathleft]{an}
\usepackage{graphicx}
\usepackage{times}
\begin{document}

\Pagespan{789}{}
\Yearpublication{2006}%
\Yearsubmission{2005}%
\Month{11}%
\Volume{999}%
\Issue{88}%

\title{Hot subdwarfs $-$ Small stars marking important events in stellar evolution}

\author{S. Geier\inst{1}\fnmsep\thanks{Corresponding author:
  \email{sgeier@eso.org}\newline}}
\titlerunning{Hot subdwarfs}
\authorrunning{S. Geier}
\institute{European Southern Observatory, Karl-Schwarzschild-Str.~2, 85748 Garching, Germany\\
Dr.~Karl~Remeis-Observatory \& ECAP, Astronomical Institute, Friedrich-Alexander University Erlangen-Nuremberg, Sternwartstr.~7, D 96049 Bamberg, Germany}
\received{}
\accepted{}
\publonline{}

\keywords{binaries: spectroscopic -- subdwarfs}

\abstract{Hot subdwarfs are considered to be the compact helium cores of red giants, which lost almost their entire hydrogen envelope. What causes this enormous mass loss is still unclear. Binary interactions are invoked and a significant fraction of the hot subdwarf population is indeed found in close binaries. In a large project we search for the close binary sdBs with the most and the least massive companions. Significantly enhancing the known sample of close binary sdBs we performed the first comprehensive study of this population. Triggered by the discovery of two sdB binaries with close brown dwarf companions in the course of this project, we were able to show that the interaction of stars with substellar companions is an important channel to form sdB stars. Finally, we discovered a unique and very compact binary system consisting of an sdB and a massive white dwarf, which qualifies as progenitor candidate for a supernova type Ia. In addition to that, we could connect those explosions to the class of hypervelocity hot subdwarf stars, which we consider as the surviving companions of such events. Being the stripped cores of red giants, hot subdwarfs turned out to be important markers of peculiar events in stellar evolution ranging all the way from star-planet interactions to the progenitors of stellar explosions used to measure the expansion of our Universe.}

\maketitle

\section{Introduction}

Looking into standard astronomy textbooks, stellar evolution appears to be quite well understood and based on established theories. That our Sun will expand to become a red giant and will eventually end her life cooling down as white dwarf tends to become common knowledge. However, this is in part a misconception, because certain phases of stellar evolution are not only less elaborated in detail, but are in fact hardly understood at all. The problem becomes even more severe, as soon as stars evolve in binary system, which is by no means an exception but rather the rule.

Mass transfer, either stable or unstable, can alter stellar evolution and form objects, which cannot be understood in a different way. Especially in the late stages of stellar evolution, where compact objects like white dwarfs and their direct precursors in very close binaries are involved, the underlying physical mechanisms can only be parametrised in a qualitative way. Among those mysterious objects, the hot subdwarf stars (sdO/Bs) stick out, because they constitute a prominent population of faint blue stars at high Galactic latitudes. With masses around $0.5\,M_{\rm \odot}$ and radii between $0.1\,R_{\rm \odot}$ and $0.3\,R_{\rm \odot}$ they are much smaller and of lower mass than hot main sequence stars of similar spectral types. Hot subdwarfs have been studied extensively for several reasons: They are common enough to account for the UV excess observed in early-type galaxies (O'Connell \cite{oconnell99}). Pulsating sdB stars became an important tool for asteroseismology (Charpinet et al. \cite{charpinet10}). The atmospheres of sdO/Bs show chemical peculiarities (Geier \cite{geier13}). SdB stars in close binaries qualify as type Ia supernova (SN\,Ia) progenitors (Maxted et al. \cite{maxted00}). And substellar companions like brown dwarfs and planets around hot subdwarfs have been discovered as well (Silvotti et al. \cite{silvotti07}). The state-of-the-art in hot subdwarf research has been reviewed by (Heber \cite{heber09}).

\begin{figure*}[t!]
\begin{center}
        \resizebox{10.5cm}{!}{\includegraphics{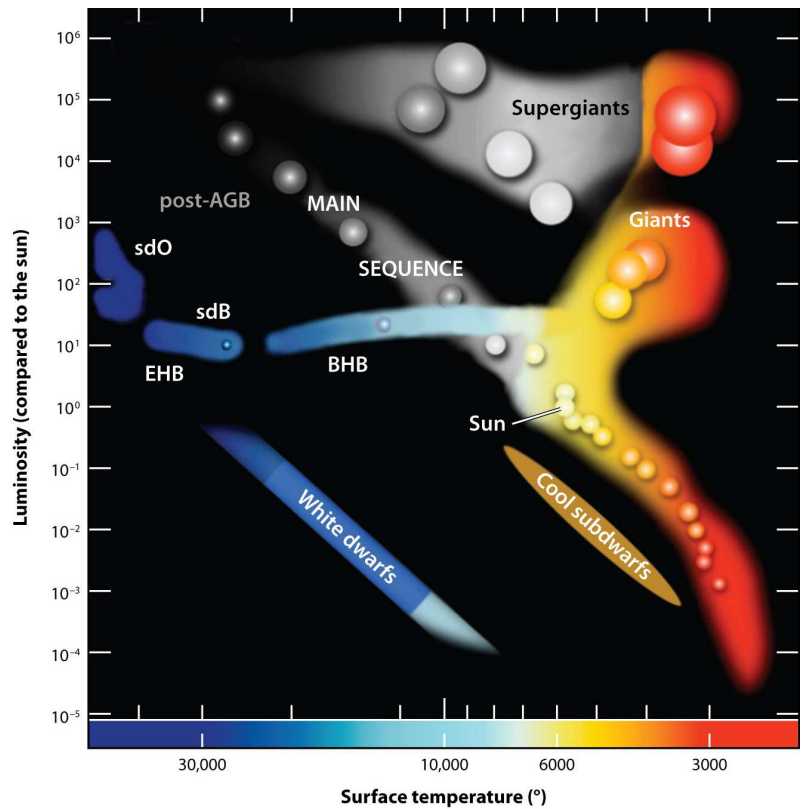}}
\end{center}
\caption{Hertzsprung-Russell diagram. The location of the extreme horizontal branch, where the sdO/Bs are located, is indicated (Heber \cite{heber09}).}
\label{hrd}
\end{figure*}

Main sequence stars burn hydrogen in their cores until this nuclear fuel reservoir is exhausted. In the next stage of evolution the stars expand and become red giants. This expansion stops as soon as helium burning starts in the red-giant cores. Depending on the amount of hydrogen envelope left after the red-giant phase, the stars in this phase occupy a region of roughly constant luminosity, but quite diverse temperatures in the Hertzsprung-Russell diagram, which is called the horizontal branch (HB, see Fig.~\ref{hrd}).
Subluminous B stars (sdB) have been identified as extreme horizontal branch (EHB) stars (Heber \cite{heber86}); i.e. they are core helium-burning stars with very thin hydrogen envelopes and therefore high temperatures. Unlike normal HB stars, which reascend the giant branch, EHB stars evolve directly to the white-dwarf cooling sequence. While the sdB stars form a homogeneous spectroscopic class, a large variety of spectra is observed among their hotter siblings, the subluminous O stars (for a detailed classification scheme of hot subdwarfs see Drilling et al. \cite{drilling13}). 

The formation of hot subdwarf stars in general is still unclear (Geier \cite{geier12a}). SdB stars can only be formed, if the progenitor loses its envelope almost entirely right at the tip of the red giant branch. While single-star scenarios are discussed, the focus shifted to binary evolution, when systematic surveys for radial velocity (RV) variable stars revealed that a large fraction of the sdB stars ($40-70\,\%$) are members of close binaries with orbital periods ranging from $0.05\,{\rm d}$ to $30\,{\rm d}$ (e.g. Maxted et al. \cite{maxted01}). While such close binaries are formed most likely after a common envelope phase, where the companion becomes completely immersed in the red-giant envelopment, stable mass transfer to a main sequence companion and the merger of two helium white dwarfs have been proposed as possible formation channels as well (Han et al. \cite{han02,han03} and references therein).

Most companions have been identified in the shortest period systems. Amongst them white dwarfs (WDs) prevail, but main sequence stars of low mass are also quite common. However, the mass range of known sdB companions widened significantly in the last couple of years. A planetary companion to a pulsating sdB star has been discovered from sinusoidal variations of its pulsation frequencies (Silvotti et al. \cite{silvotti07}). More such substellar companions in wide orbits with periods of a few hundred days have been discovered in a similar way orbiting eclipsing sdB binaries (e.g. Beuermann et al. \cite{beuermann12}). The discovery of potential close earth-size planets (Charpinet et al. \cite{charpinet11}) on the other hand showed that such objects might play a role for the formation of sdBs as well. At the other end of the mass scale, massive and compact companions like massive white dwarfs, candidate neutron stars or even black holes have been found as well (Geier et al. \cite{geier10}; Mereghetti et al. \cite{mereghetti11}). And more recently, the long sought sdBs with main-sequence companions in wide binary systems have been discovered (Vos et al. \cite{vos12,vos13}; Barlow et al. \cite{barlow13b}).

\begin{figure*}[t!]
\begin{center}
  \includegraphics[width=11.5cm]{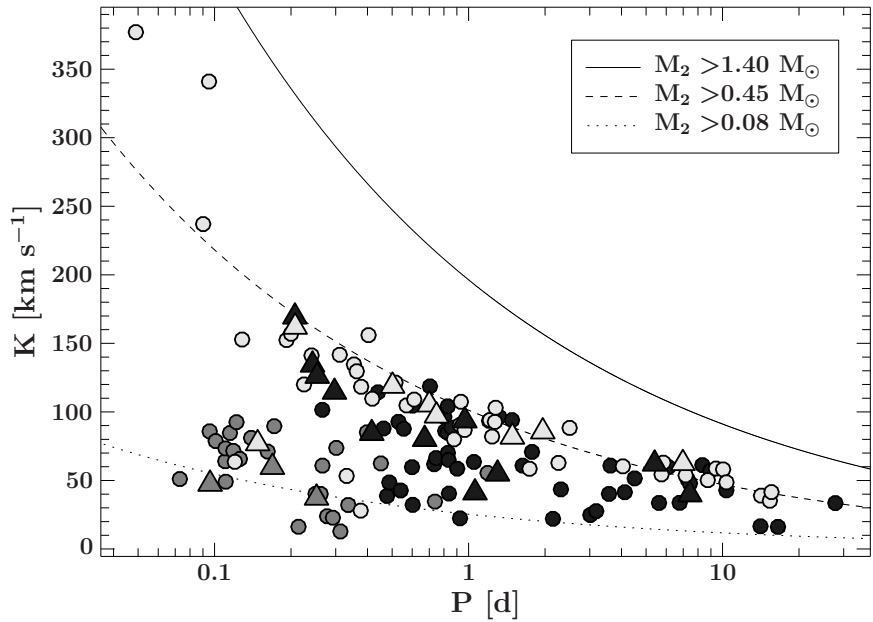}
  \caption{The RV semiamplitudes of all known sdB binaries with spectroscopic solutions plotted against their orbital periods (open symbols). The dashed, dotted and solid lines mark the regions to the right where the minimum companion masses derived from the binary mass function (assuming $0.47\,{\rm M_{\odot}}$ for the sdBs) exceed $0.08\,{\rm M_{\odot}}$, $0.45\,{\rm M_{\odot}}$, and $1.40\,{\rm M_{\odot}}$. The white symbols mark sdBs with known WD companions, the grey symbols sdBs with low-mass M-star or brown dwarf companions and the black symbols sdBs with unknown companion type (for details see Kupfer et al. \cite{kupfer15}).}
  \label{fig:binaries}
\end{center}
\end{figure*}

The formation of the subclass of helium-rich sdOs (He-sdOs) is even more enigmatic than the one of hydrogen-rich sdB stars. Most He-sdOs are concentrated at a very small region in the HRD, slightly blueward of the EHB (Str\"oer et al. \cite{stroer07}; N\'emeth et al. \cite{nemeth12}) and the population of He-sdOs observed so far seems to consist mostly of single stars (Napiwotzki \cite{napiwotzki08}). A way of forming such objects and explaining their helium-rich composition might be the merger of two helium white dwarfs (Webbink \cite{webbink84}; Iben \& Tutukov \cite{iben84}) or the delayed helium flash of a WD (Lanz et al. \cite{lanz04}). 

\section{Close hot subdwarf binaries}

To form close sdB binaries with periods down to a few hours and separations down to less than the radius of the Sun, common envelope (CE) ejection is the only likely channel. If two main sequence stars evolve in a binary system, the more massive one will evolve faster and become a red giant. Unstable mass transfer from the red giant to the companion will then lead to a CE phase. Due to the friction and gravitational drag in this phase the two stellar cores lose orbital energy and angular momentum, which leads to a shrinkage of the orbit. This energy is deposited in the envelope which will finally be ejected. 

If the core reaches the mass required for the core-helium flash before the envelope is lost, a binary consisting of a core-helium burning sdB star and a main sequence companion is formed. In another possible scenario the more massive star evolves to become a WD either through a CE phase or stable mass transfer onto the less massive companion. After that the less massive star evolves to become a red giant. Unstable mass transfer will lead to a CE and once the envelope is ejected, the red giant remnant starts burning helium and a system consisting of an sdB and a WD companion is formed (Han et al. \cite{han02,han03}). The details of the CE phase are only poorly understood (see Ivanova et al. \cite{ivanova13} and references therein) and observations of significant samples of post-CE binaries are necessary to understand this important process. 

\begin{figure*}[t!]
\begin{center}
	\resizebox{8.5cm}{!}{\includegraphics{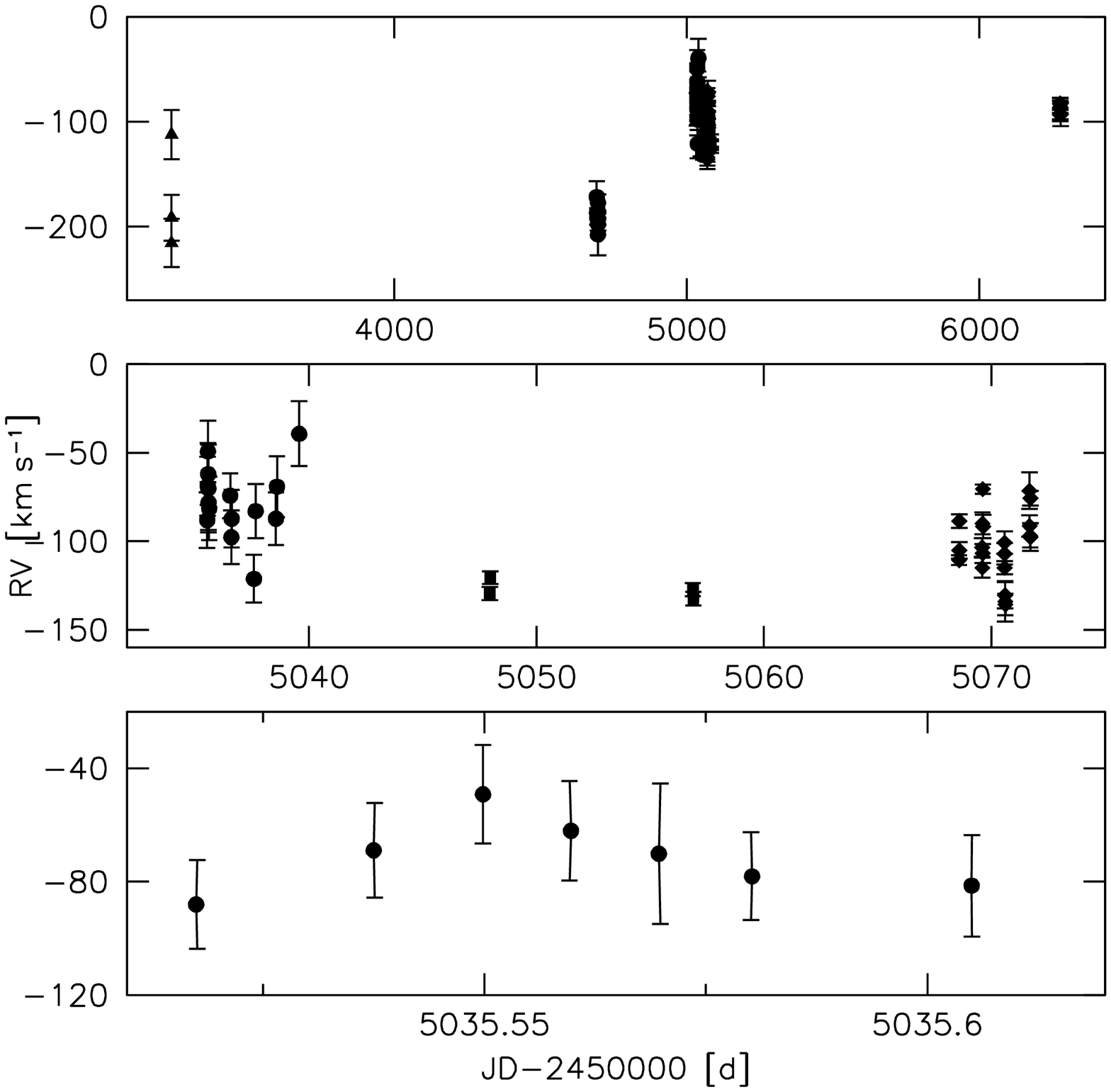}}
	\resizebox{8.5cm}{!}{\includegraphics{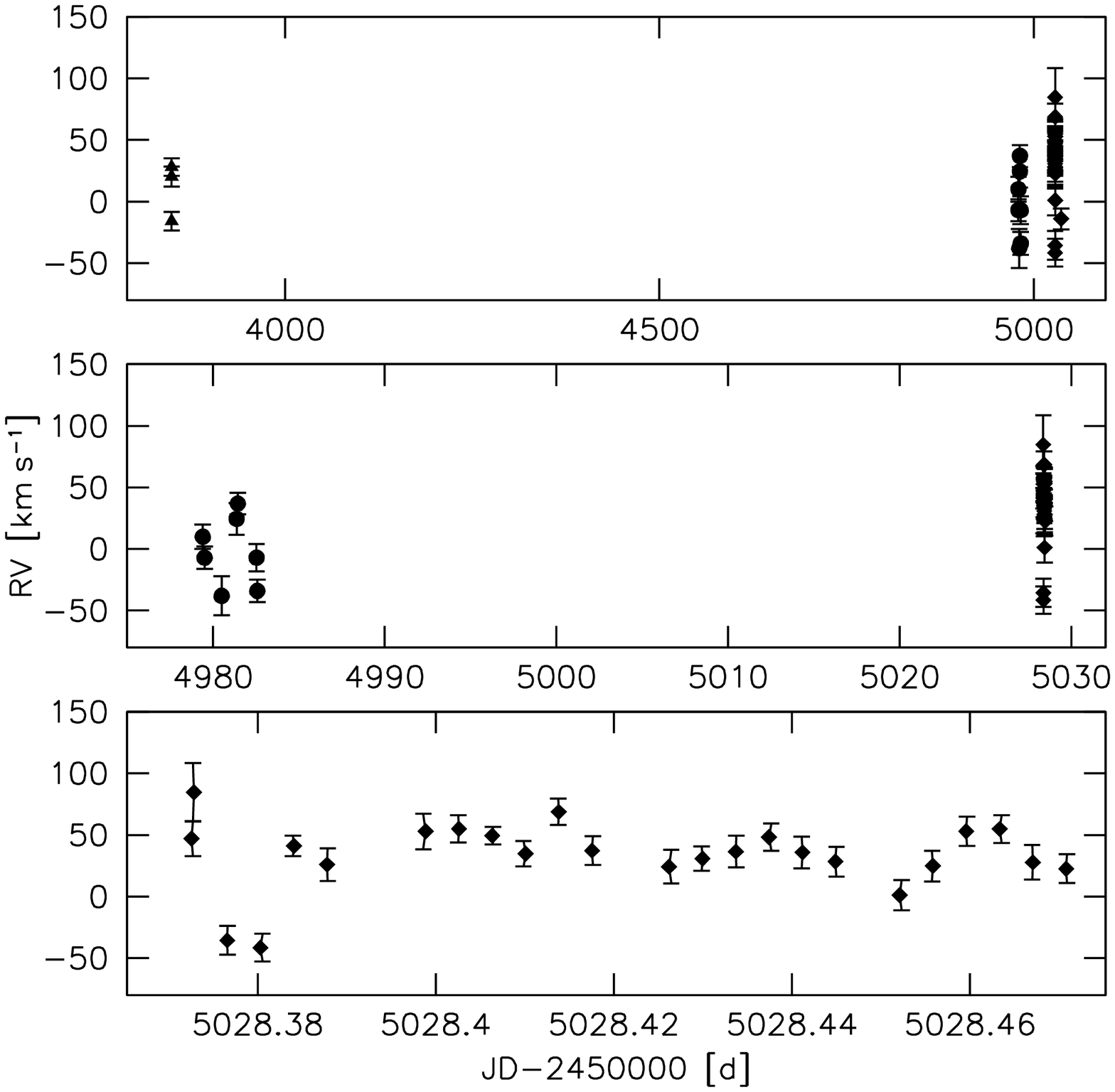}}
\end{center}
\caption{Radial velocities of the He-sdOs J232757.46+483755.2 (left panels) and J141549.05+111213.9 (right panels) against Julian date (Geier et al. \cite{geier15b}). Significant variations are present on timescales of years (upper panels), days (middle panels) and hours (lower panels).}
\label{fig:hesdo}
\end{figure*}

It is difficult to determine the nature of the close companions in sdB binaries, because they are single-lined system, where the sdB is the only component visible in the spectra. The lower mass limits derived from the binary mass functions are in general compatible with late main sequence stars of spectral type M or compact objects like white dwarfs. Only in some cases (e.g. eclipsing systems) it is possible to distinguish between those two options. 

We found, that a sizeable fraction of the sdB binary population might harbour massive compact companions, i.e. massive white dwarfs, neutron stars or even black holes (Geier et al. \cite{geier10}). The existence of such systems is actually predicted by binary evolution theory (Podsiadlowski et al. \cite{podsi02}; Pfahl et al. \cite{pfahl03}; Yungelson et al. \cite{yungelson05}; Nelemans \cite{nelemans10}). The formation channel includes two phases of unstable mass transfer and one supernova explosion. The fraction of those systems is consistently predicted to be about $1-2\%$. We determined high-precision projected rotational velocities and surface gravities of 40 sdB stars by assuming that the rotation of the sdB primary is tidally locked to its orbit. However, the synchronisation timescales as well as the sdB ages are quite uncertain. Hence, there was a need to determine companion masses for sdB stars without having to call for synchronisation.   

Motivated by this issue we carry out a radial velocity survey (Massive Unseen Companions to Hot Faint Underluminous Stars from SDSS\footnote{Sloan Digital Sky Survey}, MUCHFUSS) to find sdBs with compact companions like massive white dwarfs, neutron stars or black holes (Geier et al. \cite{geier11b}), which should show very high RV variability. We used the SDSS spectroscopic database as the starting point for our survey and obtained follow-up observations of the stars that showed the most significant RV shifts. Conducting more than $200$ nights of spectroscopic and photometric follow-up, we constrained the orbits and companion types of $30$ newly discovered sdB binaries up to now (Geier et al. \cite{geier11c,geier11d,geier13b,geier14}; \O stensen et al. \cite{oestensen13}; Schaffenroth et al. \cite{schaffenroth14b}; Kupfer et al. \cite{kupfer15}). 

Combining our new discoveries with the known close sdB binaries we performed the first comprehensive study of this population (see Fig.~\ref{fig:binaries}, Kupfer et al. \cite{kupfer15}). The minimum companion mass distribution of this sample of 142 solved close sdB binaries is bimodal. One peak around $0.1\,M_{\rm \odot}$ corresponds to the low-mass main sequence and substellar companions. The other peak around $0.4\,M_{\rm \odot}$ corresponds to the white dwarf companions. 

The derived masses for the white dwarf companions are significantly lower than the average mass for single carbon-oxygen white dwarfs, which might be an indication, that those objects are actually helium white dwarfs. We compared the sample to the population of extremely low-mass helium white dwarf binaries as well as short-period white dwarfs with main sequence companions. Both samples show a significantly different companion mass distribution indicating either different selection effects or different evolutionary paths. The sdB binaries with the shortest periods are predicted to evolve to become cataclysmic variables, stable AM\,CVn type binaries or to merge and form massive C/O WDs, RCrB stars or even to explode as SN\,Ia. 

\begin{figure*}[t!]
\begin{center}
	\resizebox{8.5cm}{!}{\includegraphics{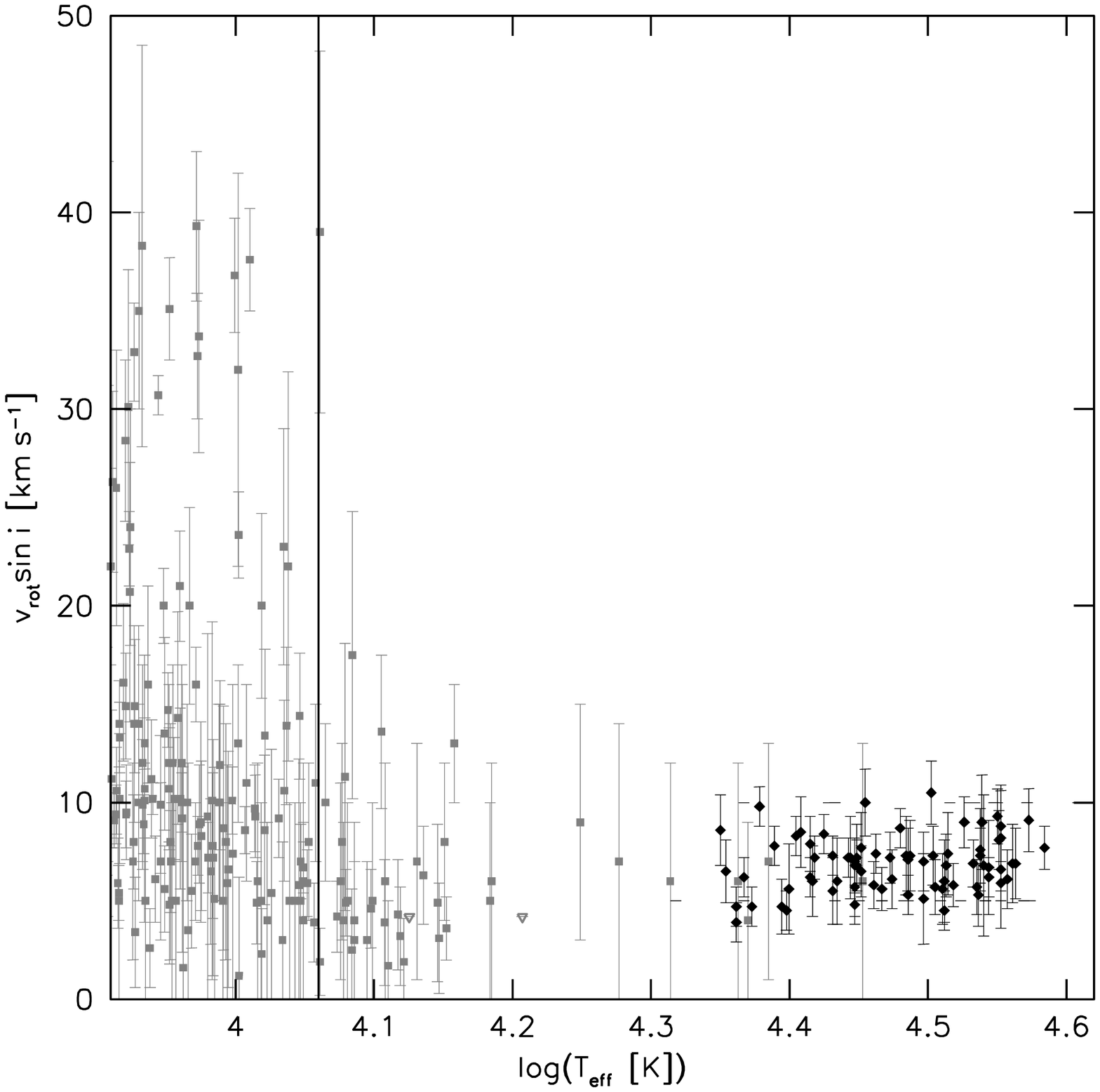}}
	\resizebox{8.5cm}{!}{\includegraphics{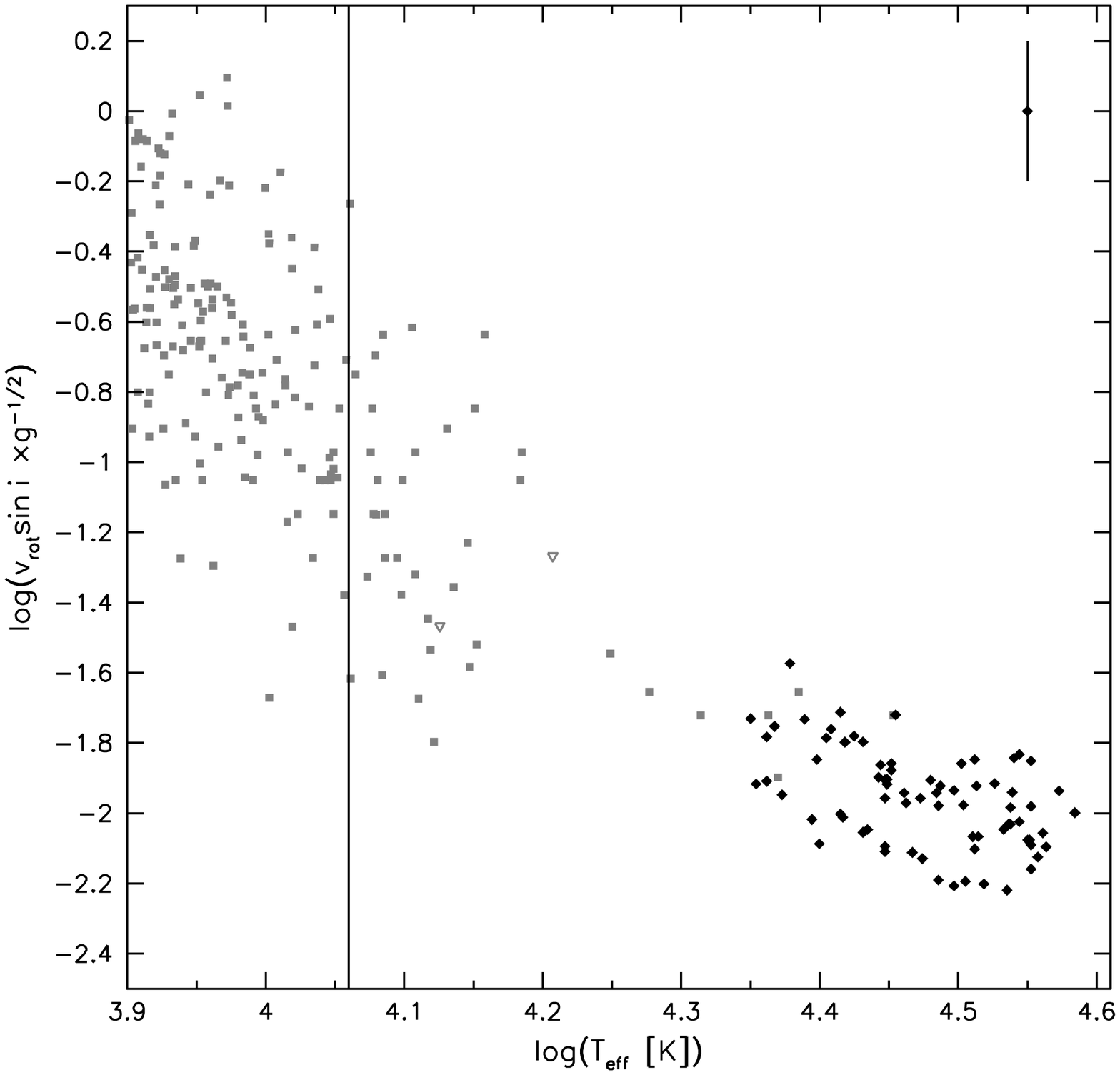}}
	\caption{{\it Left panel:} Projected rotational velocity plotted against effective temperature. The grey squares mark BHB and some sdB stars from literature. The black diamonds mark the sdBs from our sample. The vertical line marks the jump temperature of $11\,500\,{\rm K}$. {\it Right panel:} ${v_{\rm rot}\sin\,i}\times g^{-1/2}$ (proportional to the angular momentum) plotted against effective temperature for the same objects (Geier \& Heber \cite{geier12}).}
	\label{fig:vsiniteff}
\end{center}
\end{figure*}

We also published a catalogue with $1914$ radial velocity measurements of the $177$ RV variable hot subluminous stars we selected from SDSS Data Release 7. So far, we did not find an sdB binary with a compact companion exceeding $1.0\,M_{\rm \odot}$. Based on this non-detection we constrain the fraction of close massive compact companions in our sample to be smaller than $\sim1.1\%$, which is already close to the theoretical predictions. However, the sample might still contain such binaries with moderate RV-shifts and periods exceeding $\sim8\,{\rm d}$. 

Surprisingly, irregular RV variations of unknown origin with amplitudes of up to $\sim180\,{\rm km\,s^{-1}}$ on timescales of years, days and even hours have been detected in some He-sdO stars (see Fig.~\ref{fig:hesdo}). They might be connected to irregular photometric variations in some cases. Variable magnetic fields might be responsible, but no strong conclusions can be drawn yet (Geier et al. \cite{geier15b}). Another very interesting byproduct of our survey is the discovery of new classes of RV variable stars in the Galactic halo. Among them the first RV variable blue horizontal branch (BHB) stars, three candidate runaway main-sequence type B binaries (Geier et al. \cite{geier15c}) and some rare hydrogen- and helium-rich post-AGB stars (Reindl et al. \cite{reindl15}). 

\section{Substellar companions}

Soker (\cite{soker98}) suggested that substellar objects like brown dwarfs and planets may also be swallowed by their host star and that common envelope ejection could form hot subdwarfs. Substellar objects with masses higher than about $10\,M_{\rm J}$ were predicted to survive the common envelope phase and end up in a close orbit around the stellar remnant, while planets with lower masses would entirely evaporate or merge with the stellar core. The stellar remnant is predicted to lose most of its envelope and evolve towards the EHB. A similar scenario has been proposed to explain the formation of apparently single low-mass white dwarfs (Nelemans \& Tauris \cite{nelemans98}). The discovery of a brown dwarf in close orbit around such a white dwarf supported this scenario and showed that substellar companions can influence the outcome of stellar evolution (Maxted et al.  \cite{maxted06}). 

The fact that substellar companions in wide orbits around sdBs seem to be common suggests that similar objects closer to their host stars might exist as well (Silvotti et al. \cite{silvotti07}; Beuermann et al. \cite{beuermann12}). Possible signatures of earth-sized planets closely orbiting two pulsating sdB stars have been found in high-precision Kepler light curves (Charpinet et al. \cite{charpinet11}; Silvotti et al. \cite{silvotti14}). These findings indicate that planets and brown dwarf companions can survive common envelope phases. The two earth-size planets reported by Charpinet et al. (\cite{charpinet11}) have been explained either as the stripped cores of previously more massive planets or as the tidally disrupted core fragments of one massive planet (Bear \& Soker \cite{bear12}).

\begin{figure*}[t!]
\begin{center}
        \resizebox{8.5cm}{!}{\includegraphics{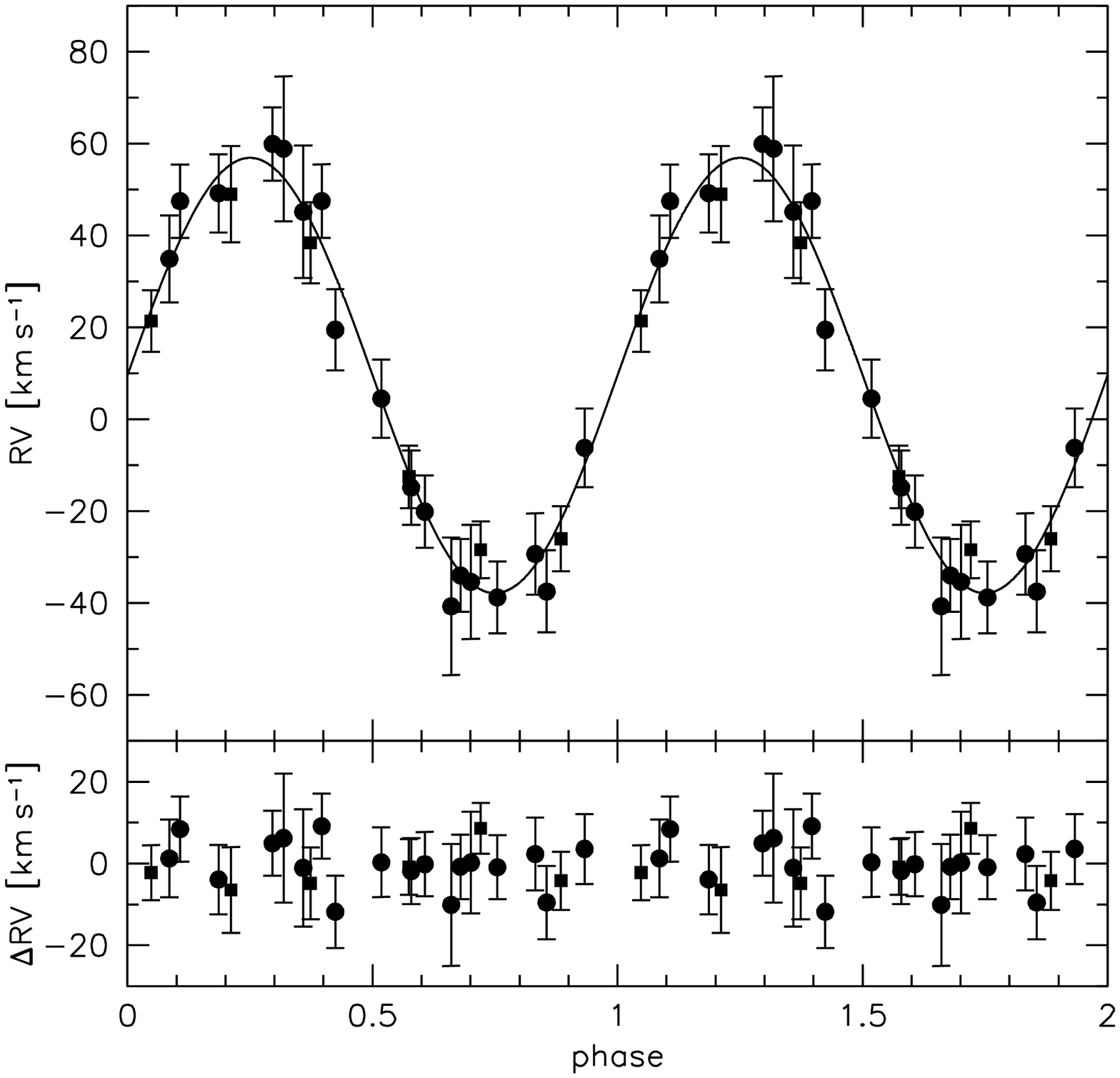}}
	\resizebox{8.5cm}{!}{\includegraphics{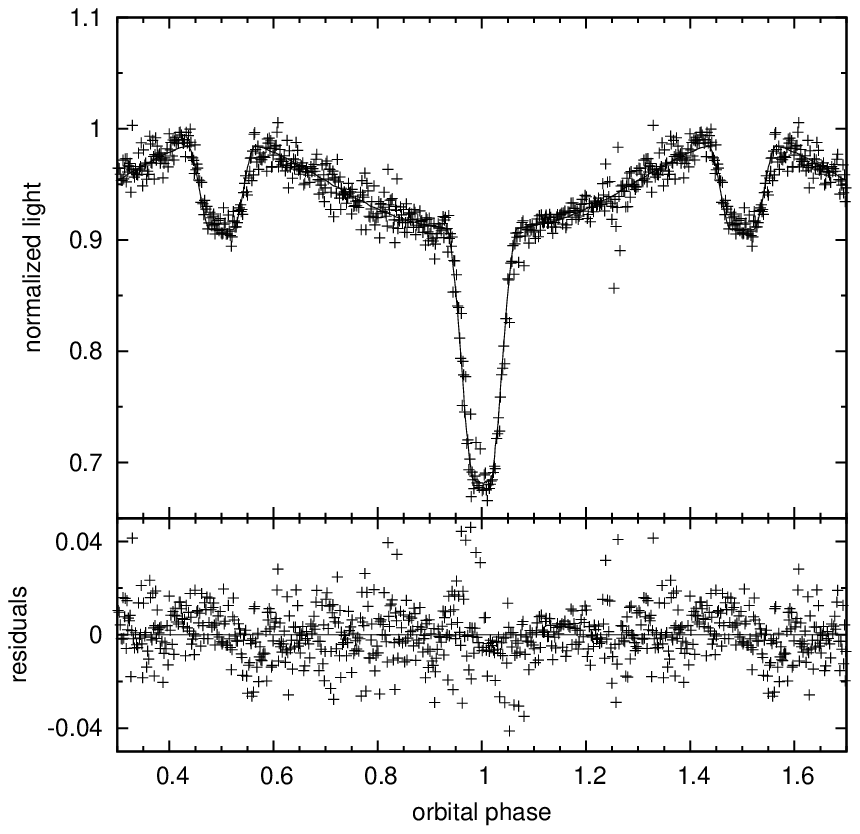}}
\end{center}
\caption{{\it Left panel:} Radial velocity plotted against orbital phase of the sdB+BD binary SDSS\,J082053+000843. {\it Right panel:} Phased R-band light curve of J08205+0008. A model is overplotted as solid line (Geier et al. \cite{geier11e}).}
\label{fig:0820}
\end{figure*}

\begin{figure*}[t!]
\centering
	\resizebox{8.5cm}{!}{\includegraphics{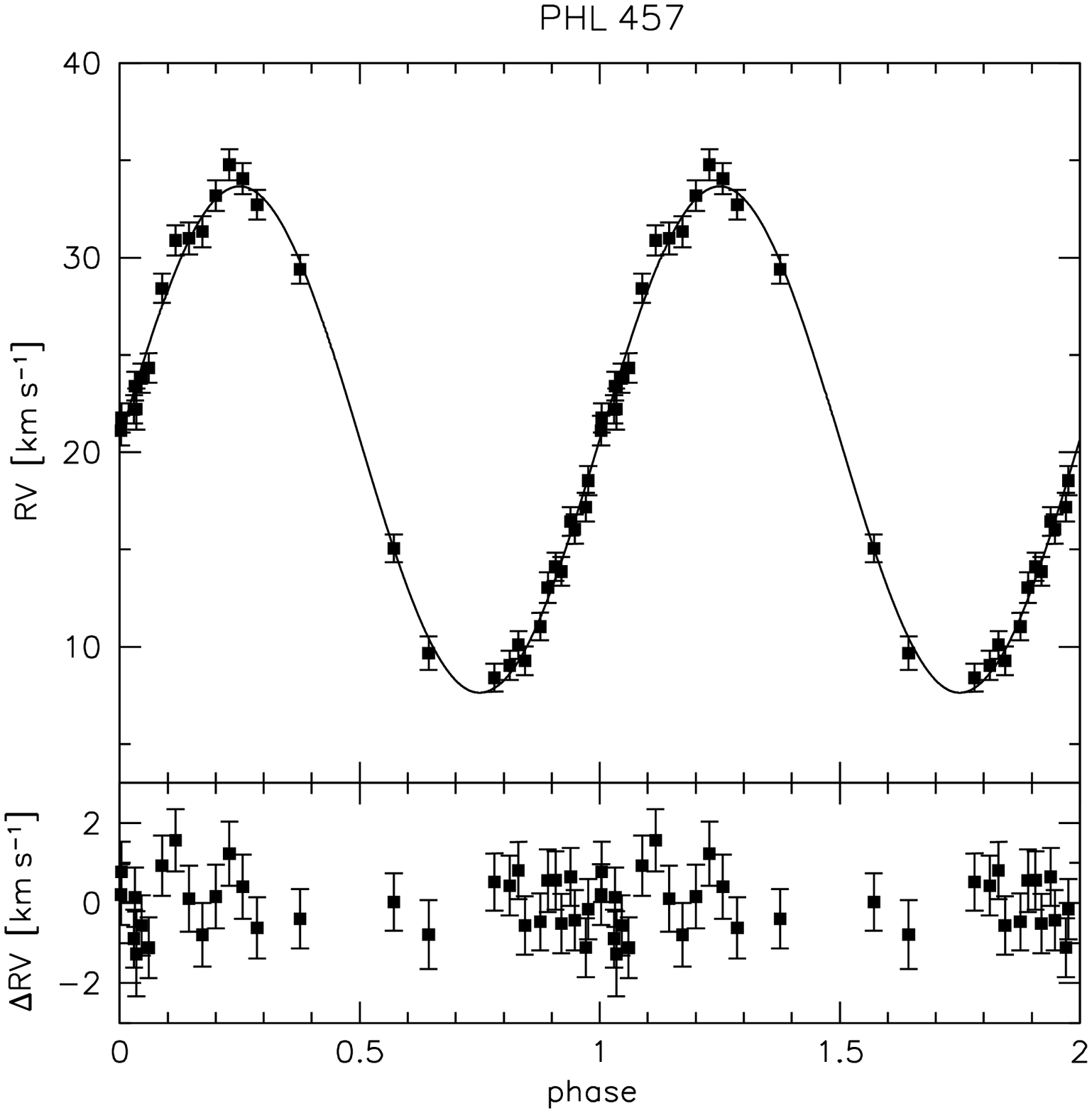}}
        \resizebox{8.5cm}{!}{\includegraphics{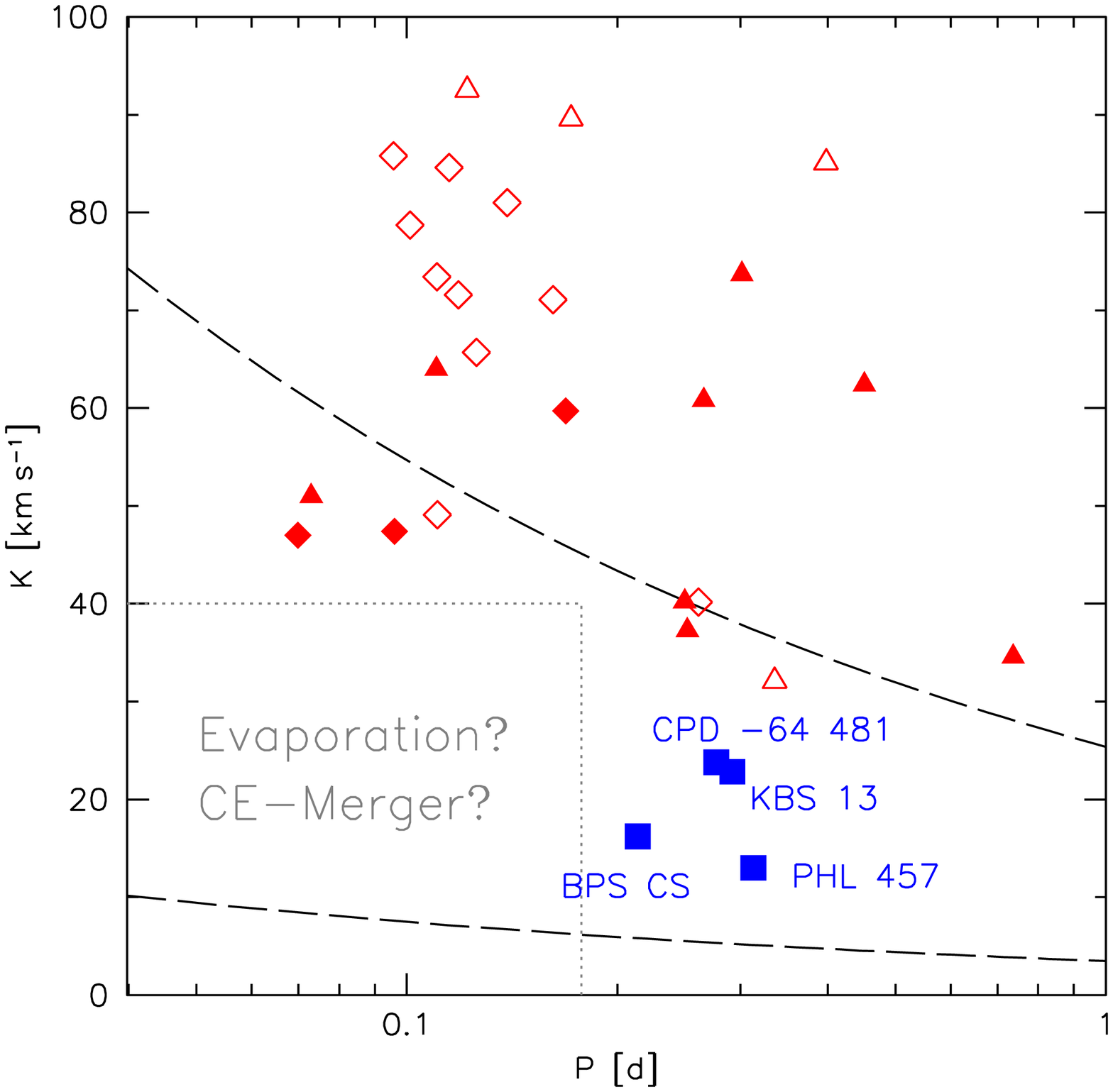}}
\caption{{\it Left panel:} Radial velocity plotted against orbital phase for PHL\,457, the sdB binary with the smallest confirmed RV variation. {\it Right panel:} The RV semiamplitudes of all known sdB binaries with reflection effects and spectroscopic solutions plotted against their orbital periods (Schaffenroth et al. \cite{schaffenroth14a}). Diamonds mark eclipsing sdB binaries where the companion masses are well constrained, triangles systems without eclipses, where only lower limits can be derived for the companion masses. Squares mark candidate sdB+BD systems. Open symbols mark systems that have been discovered based on photometry, filled symbols systems discovered based on spectroscopy. The dashed lines mark the regions to the right where the minimum companion masses derived from the binary mass function (assuming $0.47\,M_{\rm \odot}$ for the sdBs) exceed $0.01\,M_{\rm \odot}$ (lower curve) and $0.08\,M_{\rm \odot}$ (upper curve).}
\label{fig:phl}
\end{figure*}

Based on high-resolution spectra we discovered a sinusoidal variation with a very small amplitude of less than $3\,{\rm km\,s^{-1}}$ in the RV-curve of the bright sdOB HD\,149382 and concluded that it might be orbited by a planetary companion with a period of $2.39\,{\rm d}$ (Geier et al. \cite{geier09}). Although this early result could not be confirmed by other groups (Norris et al. \cite{norris11}), we soon found much better candidates. The selection criteria of the MUCHFUSS project not only single out massive companions, but also companions with very low masses in extremely short orbits. Initially, we discovered two short-period systems ($0.069\,{\rm d}$ and $0.095\,{\rm d}$) and time-resolved photometry revealed eclipses, which allowed us to constrain the companion masses. The companion mass of the eclipsing sdB binary SDSS\,J082053+000843 ($0.045-0.068\,M_{\rm \odot}$) turned out to be lower than the hydrogen-burning limit ($0.07-0.08\,M_{\rm \odot}$), which separates stars from brown dwarfs. This binary therefore hosts a brown dwarf companion and was the first such system discovered (see Fig.~\ref{fig:0820}, \cite{geier11e}). The sdB in the very similar eclipsing binary system SDSS\,J162256+473051 is also orbited by a brown dwarf companion (Schaffenroth et al. \cite{schaffenroth14b}). 

These results, which showed that brown dwarfs can trigger and survive a common envelope ejection, provided the best evidence so far that substellar companions play an important role in the formation of sdB stars. We therefore searched for more sdB binaries with low-mass companions and found two close reflection effect binaries with grazing eclipses and low-mass stellar companions (Schaffenroth et al. \cite{schaffenroth13,schaffenroth15}). In addition to a reflection effect, which signals the presence of a cool, low-mass stellar companion, we detected p-mode pulsations in the light curve of the sdB binary FBS\,0117+396. Only a few of the known short-period sdB pulsators are in close binary systems (\O stensen et al. \cite{oestensen13}). Furthermore, I became involved in several other studies of sdB binaries with low-mass stellar companions (Geier et al. \cite{geier11d}; Naslim et al. \cite{naslim12}; Barlow et al. \cite{barlow13a}; Telting et al. \cite{telting14}). 

\begin{figure*}[t!]
\centering
        \resizebox{8.5cm}{!}{\includegraphics{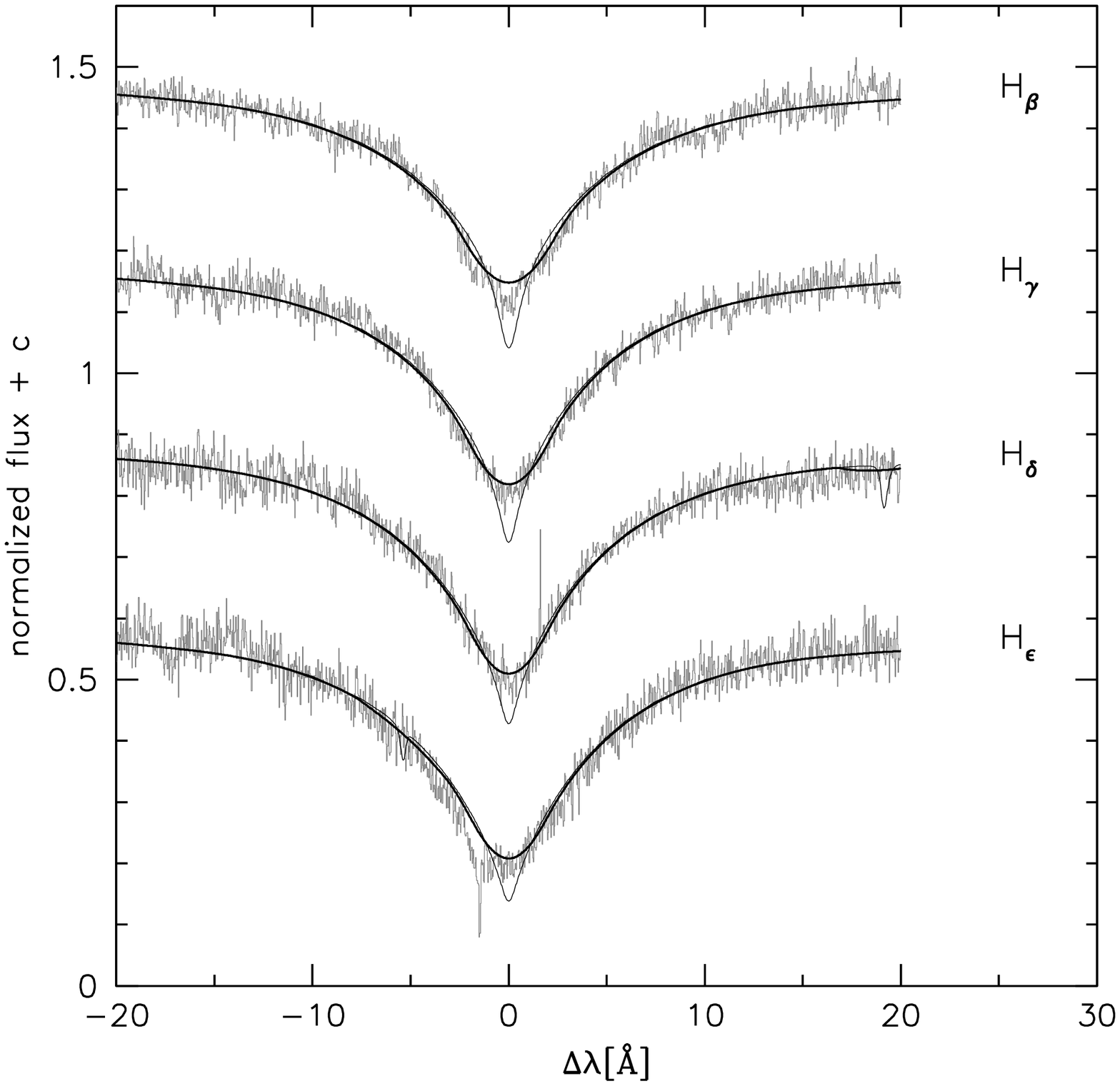}}
        \resizebox{8.5cm}{!}{\includegraphics{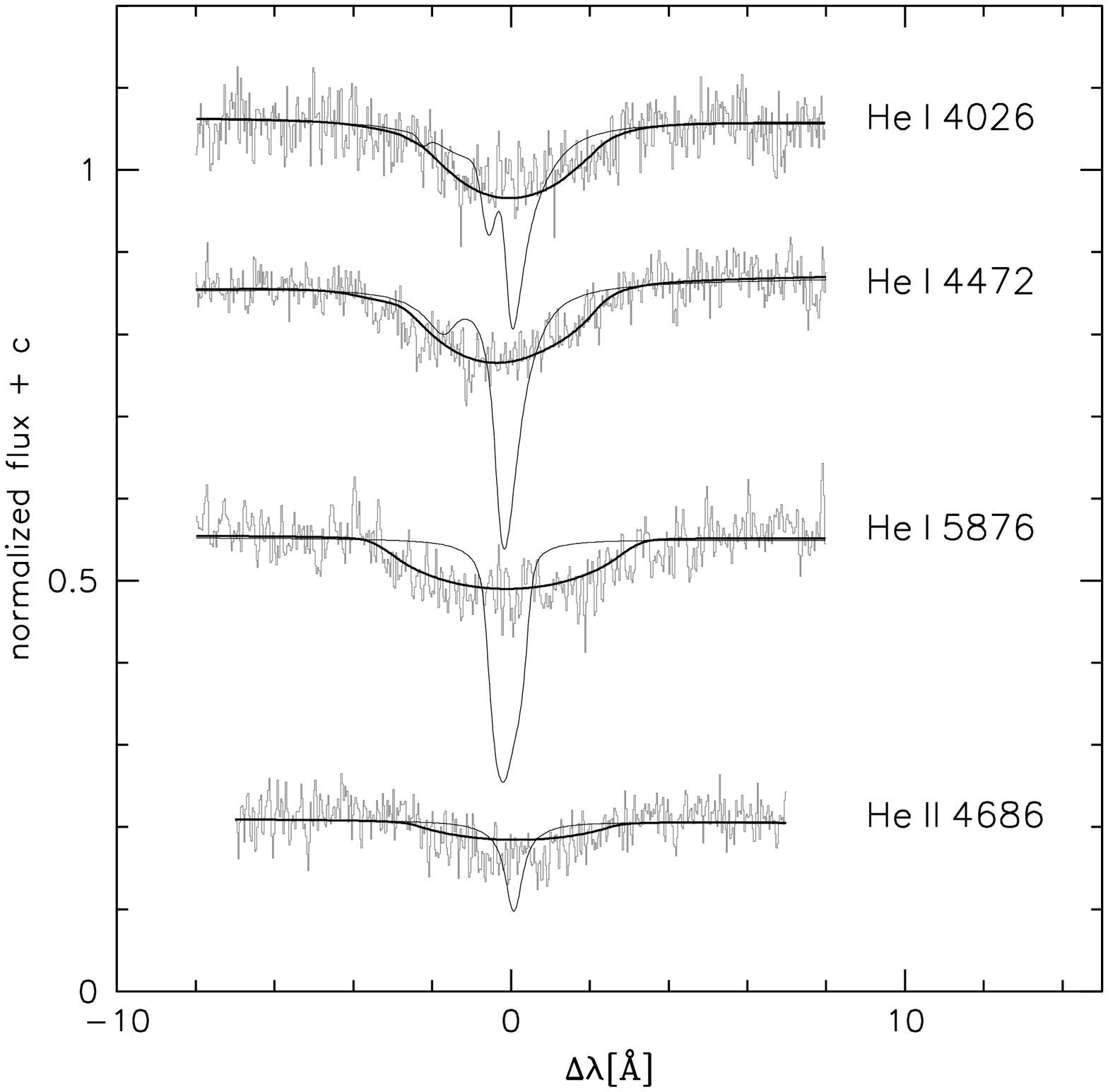}}
\caption{{\it Left panel:} Fit of synthetic models to some hydrogen Balmer lines of the sdB star EC\,22081$-$1916. The thin solid line marks models without rotational broadening, the thick solid line the best fitting model spectrum with $v_{\rm rot}\sin{i}=163\,{\rm km\,s^{-1}}$. {\it Right panel:} Fit of synthetic models to helium lines. The extreme rotational broadening of the lines is obvious (Geier et al \cite{geier11a}).}
\label{fig:ec22081}
\end{figure*}

Our goal is to increase the sample of sdB binaries with substellar companions and to check, whether there is a lower mass limit for the surviving companion as suggested by Soker (\cite{soker98}). Recently, we found two new reflection effect binaries with cool companions and very small minimum companion masses down to $0.027\,M_{\rm \odot}$ and had studied the known sample of sdBs with low-mass stellar companions (see Fig.~\ref{fig:phl} left panel, Schaffenroth et al. \cite{schaffenroth14a}). We conclude that the fraction of sdBs formed after an interaction with low-mass stars is comparable to the fraction of sdBs formed after an interaction with substellar objects. Furthermore, we do not find any binaries with orbital periods shorter than $\sim0.2\,{\rm d}$ having minimum masses below $\sim0.06\,M_{\rm \odot}$ and argue that substellar companions in this range might not survive the common envelope phase (see Fig.~\ref{fig:phl} right panel). Instead they might either merge with the core of the red giant or evaporate in its envelope as suggested by Soker (\cite{soker98}).  

Independently, we found more evidence for this scenario. Studying a sample of bright sdBs, we discovered two single stars, which are fast rotators, while the large majority of single sdBs are very slow rotators (see Fig.~\ref{fig:vsiniteff}, Geier \& Heber \cite{geier12}). Since tidal forces spin up the rotation of sdBs in close binary stars, we initially took rapid rotation as indication for close companions (Geier et al. \cite{geier10}). We found EC\,22081$-$1916 to show strongly rotationally broadened line profiles (see Fig.~\ref{fig:ec22081}, Geier at al. \cite{geier11a}). However, time-resolved spectroscopy did not show any RV variations indicative of a close companion. This lead us to the conclusion that the star was formed by the  common envelope merger of a low-mass, possibly substellar, object with a red-giant core just as proposed by Soker (\cite{soker98}) and further elaborated by Politano et al. (\cite{politano08}). Subsequently, we found the single sdB SB\,290 to behave in a similar way (Geier et al. \cite{geier13a}). 

While a common envelope merger event is predicted to form a fast rotating sdB (Politano et al. \cite{politano08}), it is much harder to prove the evaporation of a low-mass companion just before common envelope ejection, because the companion does not survive the interaction and should not have any measurable influence on the formed sdB. However, the mere existence of single sdB stars might already be an indication for such a process. Based on the very slow rotation of apparently single sdBs we showed that the merger of two He-WDs is an unlikely formation scenario for sdB stars (Geier \& Heber \cite{geier12}). The merger scenario is also not consistent with the very narrow mass distribution of sdB stars determined from asteroseismic and close binary analyses (Fontaine et al. \cite{fontaine12}). While the formation of sdBs in binaries can be explained, we are lacking such a scenario for single sdBs. 

\begin{figure*}[t!]
\centering
	\resizebox{14.5cm}{!}{\includegraphics{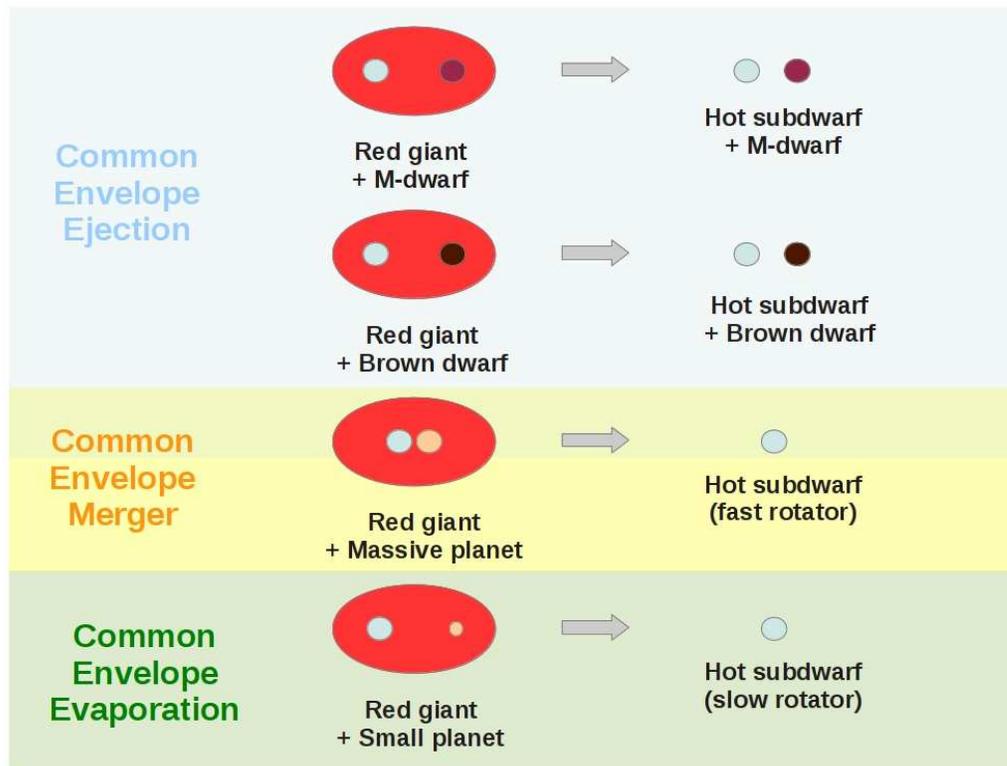}}
\caption{Schematic view of sdB formation via the interaction with low-mass stellar and substellar objects.}
\label{fig:planets}
\end{figure*}

One possible explanation might be that there are in fact no single sdBs, because the companions with the smallest masses just remained undetected so far. Because searches for RV-variability have mostly been done using medium-resolution spectra, which only allowed us to measure RV shifts higher than $\sim10\,{\rm km\,s^{-1}}$, we measured RVs from our high-resolution spectra with an accuracy down to less than $\sim1.0\,{\rm km\,s^{-1}}$ to check whether a yet undetected population of sdB binaries with small RV amplitudes (see Fig.~\ref{fig:phl}) might be present. Preliminary results show that companions in close orbits with masses down to $0.01\,M_{\rm \odot}$ can be excluded in more than half of our sample meaning that single sdBs indeed exist (Classen et al. \cite{classen11}). One possible way to explain their existence might be the interaction with a close companion, which has been evaporated. A schematic view of possible interactions between red giants and low-mass stellar or substellar objects resulting in the formation of sdBs is shown in Fig.~\ref{fig:planets}.

\begin{figure*}[t!]
 \includegraphics[width=10.5cm]{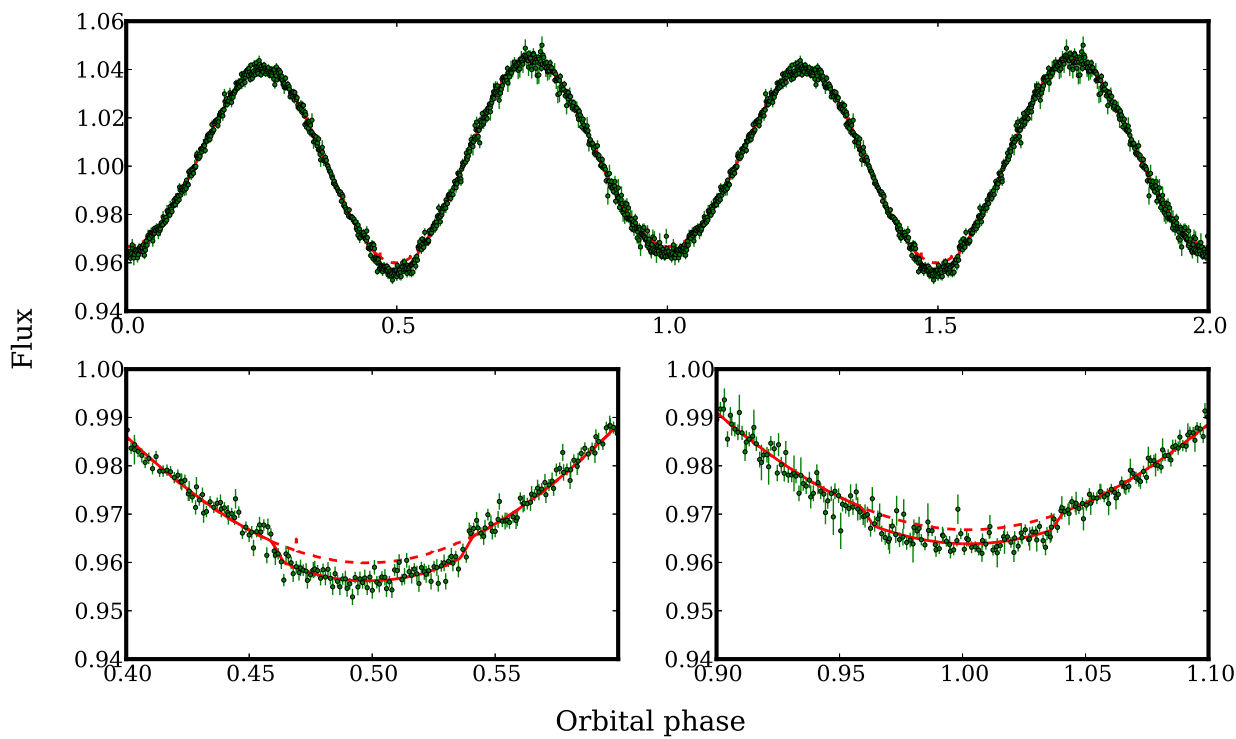}
 \caption{{\it Upper panel:} V-band light curve of CD$-$30$^\circ$11223 taken with SOAR/Goodman (green) with superimposed model (red) plotted twice against orbital phase for better visualisation. The dashed red curve marks the same model without transits and eclipses. The sinusoidal variation is caused by the ellipsoidal deformation of the hot subdwarf due to the tidal drag of the compact white dwarf. The difference in amplitude between phase $0.25$ and $0.75$ originates from the relativistic Doppler beaming effect, which is usually not detectable with ground-based telescopes. {\it Lower panels:} Close-up on the transit of the WD in front of the sdB (left). The WD companion is comparable in size to Earth. The detection of transits is essential to determine the fundamental parameters of the binary using model light curves. It is even possible to detect the eclipse of the WD by the sdB (right), which tells us that the WD is still young and therefore hot enough to contribute significant flux (Geier et al. \cite{geier13b}).}
 \label{cdm30_lc}
\end{figure*}

\section{Supernova type Ia progenitors and ejected donor remnants}

Using SN Ia as cosmological standard candles provided the first and still most direct evidence for the accelerated expansion of the universe (Nobel prize 2011, Riess et al. \cite{riess98}; Perlmutter et al. \cite{perlmutter99}). However, the accuracy of this method is limited, because the progenitors of SN Ia explosions are still unknown. There is consensus that the observed features of SN\,Ia can only be conclusively explained by the thermonuclear explosion of a white dwarf consisting of carbon and oxygen. To reach the critical mass for the ignition, matter must be transferred from a close companion star. Only if the progenitor systems of SN Ia are properly understood, we can hope to use these tools to gain a deeper insight into the nature of the dark energy and the evolution of our universe (see Wang \& Han \cite{wang12} for a review). 

Close subdwarf binaries with massive WD companions turned out to be candidates for SN Ia progeni\-tors because those systems shrink further due to the emission of gravitational waves and merge. One of the best known candidate systems for this double-degenerate merger scenario is the sdB+WD binary KPD\,1930$+$2752 (Maxted et al. \cite{maxted00}; Geier et al. \cite{geier07}). Another possible channel for SN\,Ia is the single-degenerate scenario where a massive white dwarf accretes matter from a close companion until it reaches the critical mass and explodes. Mereghetti et al. (\cite{mereghetti09}) showed that in the X-ray binary HD\,49798 a massive ($>1.2\,M_{\rm \odot}$) white dwarf accretes matter from a closely orbiting subdwarf O companion.

In the course of the MUCHFUSS project we discovered the extremely close ($P=0.04987\,{\rm d}$), eclipsing binary system CD$-$30$^\circ$11223, which fulfills all requirements for the progenitor of a SN\,Ia via yet another channel, the so-called sub-Chandrasekhar double-detonation scenario (Fink et al. \cite{fink10}). In this system a white dwarf ($\sim0.8\,M_{\rm \odot}$) is orbited by a core-helium burning compact hot subdwarf star, which will start to transfer helium-rich material on short timescales. The ignition of helium burning in the newly formed envelope ($\simeq0.1\,M_{\rm \odot}$) is predicted to trigger carbon-burning in the core although the WD is less massive than the Chandrasekhar limit (see Fig.~\ref{cdm30_lc}). Furthermore, we argued that the hypervelocity sdO US\,708 is likely to be the surviving donor remnant of such an event (Geier et al. \cite{geier13b}; see also Vennes et al. \cite{vennes12}). 

This connection between SN\,Ia and surviving helium stars, that are ejected from the very close progenitor binaries at high velocities, has been predicted by theory (Justham et al. \cite{justham09}; Wang et al. \cite{wang09}). In a follow-up study of US\,708 we provided further evidence for this connection. This star is not only a fast rotator as expected, because it must have been spun-up in the close binary before. Its origin can also be traced back to the Galactic disc rather than the Galactic centre (Geier et al. \cite{geier15a}). In this way, essentially all other proposed acceleration scenarios for this star can be excluded (e.g. Hills \cite{hills88}). Measuring all velocity components of US\,708 for the first time, we find it to be the fastest unbound star in our Galaxy. If it can be shown, that high velocity sdO/B stars are indeed the ejected donor remnants, it would be a direct proof that the double-detonation scenario leads to SN\,Ia explosions. The first type of SN\,Ia progenitors would then be unambiguously identified. 

\section{Conclusion}

Since their discovery in the middle of the last century, hot subdwarf stars have been studied for quite different reasons. First they appeared in the colour-magnitude diagrams of globular clusters providing another hint that those clusters are more complicated than expected before. Bright hot subdwarfs in the field became testbeds for the modelling of hot stellar atmospheres and their chemical peculiarities still pose a challenge in this respect. Later on, different kinds of pulsations where discovered and hot subdwarfs became favourable objects to apply asteroseismic models. 

Explaining their formation posed a challenge all along and still does today. However, significant progress has been made in the last years observing large samples of hot subdwarf stars in different stellar populations and studying their properties. Hot subdwarfs turned out to be quite numerous and because of their very blue colours quite easy to spot in old and red populations. Most of them seem to be the remnants of interactions between stars or between stars and substellar or evolved objects. These properties make them key objects to study such interactions in general and with them phenomena as diverse as the interactions of stars and planets or the progenitors of supernovae type Ia. 

While some evolved stars like white dwarfs, which are much more numerous, might have experienced similar episodes of interaction, those peculiar objects will always be needles in the giant haystack of their respective unperturbed populations and therefore very difficult to find. Hot subdwarfs on the other hand seem to be formed exclusively via non-standard channels and each of them is therefore an important marker of such events in stellar evolution.

\end{document}